\documentclass[12pt,epsf,epsfig]{article}

\usepackage{epsfig}
\usepackage{epsf}
\newcommand\iso[2]{\mbox{${}^{#2}{\rm #1}$}}
\newcommand\he[1]{\iso{He}{#1}}
\newcommand\li[1]{\iso{Li}{#1}}
\newcommand\be[1]{\iso{Be}{#1}}
\newcommand\omb{\Omega_{\rm B}}
\newcommand\nnu{N_{\rm \nu,eff}}
\def\ion#1#2{#1$\;${\small\rm #2}\relax}
\newcommand\hii{\ion{H}{II}}
\def\ga{\mathrel{\raise.3ex\hbox{$>$\kern-.75em\lower1ex\hbox{$\sim$}}}}
\def\la{\mathrel{\raise.3ex\hbox{$<$\kern-.75em\lower1ex\hbox{$\sim$}}}}

\newcommand\beq{\begin{equation}}
\newcommand\eeq{\end{equation}}
\newcommand\beqar{\begin{eqnarray}}
\newcommand\eeqar{\end{eqnarray}}
\begin{document}

\begin{titlepage}
\pagestyle{empty}
\baselineskip=21pt
\vspace*{-0.6in}
\rightline{astro-ph/0302431}
\rightline{UMN--TH--2130/03}
\rightline{TPI--MINN--03/05}
\vskip 0.2in
\begin{center}
{\large{\bf Primordial Nucleosynthesis in Light of {\it WMAP}}}
\end{center}
\begin{center}
\vskip 0.2in
{{\bf Richard H. Cyburt}$^1$, {\bf Brian D. Fields}$^2$
and {\bf Keith A. Olive}$^3$}\\
\vskip 0.1in
{\it
$^1${Department of Physics, University of Illinois, Urbana, IL 61801}\\
$^2${Center for Theoretical Astrophysics,
Department of Astronomy, \\ University of Illinois, Urbana, IL 61801}\\
$^3${William I. Fine Theoretical Physics Institute, \\
University of Minnesota, Minneapolis, MN 55455, USA}}\\
\vskip 0.2in
{\bf Abstract}
\end{center}
\baselineskip=18pt \noindent
Big bang nucleosynthesis has long provided
the primary determination of the cosmic baryon density $\omb
h^2$, or equivalently the baryon-to-photon ratio, $\eta$.
Recently, data on CMB anisotropies have become
increasingly sensitive to
$\eta$.  The comparison of these two independent
measures provides
a key test for big bang cosmology.
The first release of results from
the {\it Wilkinson Microwave Anisotropy Probe} (WMAP)
marks a milestone in this test.  With the precision of WMAP,
the CMB now offers a significantly stronger constraint on
$\eta$.  
We discuss the current state of BBN theory and light element observations
(including their possible lingering systematic errors).
The resulting BBN baryon density prediction is in overall agreement
with the WMAP prediction, an important and non-trivial confirmation of 
hot big bang cosmology. Going beyond this, the powerful CMB baryometer
can be used as an input to BBN and one can accurately predict the
primordial light element abundances. By comparing these with observations
one can obtain new insight into post-BBN nucleosynthesis processes and
associated astrophysics. Finally, one can test
the possibility of nonstandard
physics at the time of BBN, now with all light elements available as
probes.  Indeed, with the WMAP precision 
$\eta$, deuterium is already beginning to rival \he4's sensitivity to
nonstandard physics, and additional D/H measurements can improve this
further.
\end{titlepage}
\baselineskip=18pt

\section{Introduction}

The primordial light element abundances are predicted accurately and
robustly by the theory of Big Bang Nucleosynthesis (BBN) \cite{bbn,sarkar},
describing the first 3 minutes of the hot early universe.  This hot big
bang model also predicts a relic photon background, produced when nuclei
recombined to form neutral atoms some 400,000 years later.  The Cosmic
Microwave Background (CMB), and its anisotropies carry key information
about the content of the universe and early structure growth.  In
particular, both BBN and the CMB are sensitive to the baryon content in
the universe and because they are governed by different physics, BBN and
the CMB can be used as independent measures of the cosmic baryon density,
$\rho_{\rm B}\propto
\omb h^2$, or equivalently the baryon-to-photon ratio, $\eta$.

The comparison of the baryon density predictions from BBN and the CMB
is a fundamental test of big bang cosmology \cite{st}, and its
underlying assumptions, which include:
a nearly homogeneous, isotropic universe,
with gravity described by General Relativity and 
microphysics described by the Standard Model of particle 
physics.\footnote{Other, somewhat more technical assumptions are
that no comoving entropy change occurs between BBN and the CMB,
and that the neutrino chemical potentials are small,
i.e., that the cosmic lepton number $n_L/n_\gamma \ll 1$.} In the standard model, we
fix the number of neutrino flavors to three, and we allow this number to
vary in order to test models beyond the standard model.
Furthermore, standard BBN relies on a network of nuclear reactions
which are taken from low energy cross section measurements.   
Any deviation from
concordance points to either unknown systematics or the need for new
physics.  Up till now, there has been tentative agreement between the
baryon density predictions from BBN and the CMB, barring the internal
tension between BBN derived limits from deuterium and \li7
observations.  With the first data release from the Wilkinson
Microwave Anisotropy Probe (WMAP), the anisotropies in the CMB have been
measured to unprecedented accuracy \cite{wmap}.  This new precision allows
for a CMB-based determination of the baryon density 
which is significantly tighter than current BBN analysis yields.  
One no longer needs to use BBN as a probe of the baryon
density.  Instead, the CMB baryon density can be used as an input for BBN,
and the light element abundance observations can be used to test particle
physics and nuclear astrophysics \cite{cfo2,kssg}.

This paper is organized as follows.  In section~\ref{sect:bbn}, we
discuss the state of affairs of primordial nucleosynthesis before
WMAP.  We then explain how the post-WMAP CMB compares with BBN in
section~\ref{sect:cmb}, and go on to constrain astrophysics
(section~\ref{sect:bbn+cmba}) and particle physics
(section~\ref{sect:bbn+cmbp}).  We conclude with a discussion of our
results and aspirations for the future.

\section{The Baryon Density from BBN (Pre-WMAP)}
\label{sect:bbn}

The baryon density (or the baryon-to-photon ratio, 
$\eta \equiv \eta_{10}/10^{10}$) is the sole
parameter in the standard model of BBN. Prior to the recent measurements
of the microwave background power spectrum, the best available method 
for
determining the baryon density of Universe was the concordance of the 
BBN
predictions and the observations of the light element abundances of D,
\he3, \he4, and
\li7. A high-confidence upper limit to the baryon density
has long been available \cite{rafs} 
from observations of local
D/H abundance determinations (giving roughly $\eta_{10} < 9.0$),
but a reliable lower bound to $\eta$, much less a precise value, has been
more elusive to obtain. Lower bounds to $\eta$ have been derived (1)  on
the basis of D + \he3 observations (using arguments based on chemical
evolution) \cite{ytsso}, (2) from early reports (now understood to be
erroneous) of high D/H in  quasar
absorption systems, and 
(3) in likelihood analyses using the combined \he4, \li7 and D/H observations
\cite{fo,lik,fior,cfo1}. The last method gives a 95 \% CL range of $5.1 <
\eta_{10} < 6.7$ with a most likely value of $\eta_{10} = 5.7$
corresponding to
$\Omega_B h^2 = 0.021$.

Observations of each of the light elements D, \he4, and \li7 can be used to
determine the value of $\eta$. Despite great progress theoretically and
observationally \cite{brb}, \he3
is not as yet a strong baryometer \cite{vofc} (but see below, \S
\ref{sect:bbn+cmba}).
Each of the light elements is observed in vastly different astrophysical
environments: D/H in high-redshift QSO absorption line systems;  \he4 in
extragalactic \hii\ regions; and \li7 in low metallicity halo stars.
Confidence in any such determination however, relies on the concordance
of the three light isotopes.
  One concern regarding the likelihood method is, in fact, the
relatively poor agreement between \he4 and \li7 on the one hand and
D on the other.  The former two taken alone indicate that the most 
likely
value for $\eta_{10}$ is 2.4, while D/H alone implies a best value of 
6.1.
This discrepancy may point to new physics, but could well be due to
underestimated systematic errors in the observations.
More weight has been given to the D/H determinations because of their
excellent agreement with the (pre-WMAP) CMB experiments.

\section{The Baryon Density from the CMB and Beyond}
\label{sect:cmb}

The power spectrum of CMB temperature anisotropies
contains a wealth of information about a host of cosmological parameters,
including  $\eta$
\cite{zss}.
In the past few years, pioneering balloon and ground-based observations have
made the first observations at multipoles $\ell \ga 200$, 
where the sensitivity to $\eta$ lies, and constraints on $\eta$
reached near the sensitivity of BBN \cite{oldcmb}.
Already, these experiments had revealed the first two
acoustic peaks in the angular power spectrum,
and hints of a third.
The improvement offered by WMAP \cite{wmap} was thus
a quantitative one:  with its all-sky coverage, high signal-to-noise,
and broad angular coverage, WMAP offers a major advance
in our understanding of the CMB and allows 
the CMB-based inference of the baryon-to-photon ratio
to reach a new level of precision.

The CMB-based baryon density 
must be extracted from 
the observed angular power spectrum of temperature anisotropies.
This process requires several assumptions.
In addition to adopting the basic
hot big bang framework, outlined above, some more specific
assumptions are required. 
These are: 
(1) gaussian random fluctuations,
(2) flat priors over the adopted range of parameters;
(3) an adiabatic primordial power spectrum of density fluctuation 
described by a single, constant spectral index,
or by an index with a constant logarithmic slope versus $k$.
The baryon density is then determined simultaneously with several other key cosmological 
parameters which include: the total matter density, the Hubble parameter, spectral index and
optical depth. 
In addition, other data sets can be adopted to further
constrain the cosmological parameters (including $\omb$).
The WMAP best fit result is for a varying spectral index,
and is \cite{wmap} 
$\omb h^2 = 0.0224 \pm 0.0009$, or
\beq
\label{eq:eta-cmb}
\eta_{\rm 10,CMB} = 6.14 \pm 0.25
\eeq
a precision of 4\%!
This estimate is the best-fit WMAP value,
which is sensitive mostly to WMAP alone (primarily the first
and second acoustic peaks) but does include
CBI \cite{cbi} and ACBAR \cite{acb} data on smaller angular scales,
and Lyman $\alpha$ forest data (and 2dF redshift survey data \cite{2df})
on large angular scales.

The various data sets, and assumptions regarding the
spectral index, all influence the ``best fit'' WMAP baryon density.
For WMAP data alone, the baryon density is
$\omb h^2 = 0.024 \pm 0.001$ for a constant
spectral index in a $\Lambda$ CDM cosmology; 
this value is about $1.6 \sigma$ above the best fit.  
The CBI and ACBAR data serve to decrease $\omb h^2$ by 
about $0.001$ units, and the Lyman $\alpha$  data
make a smaller shift, but in opposite directions depending on
the constant or running nature of the spectral index.
For the rest of the paper, 
unless stated otherwise we will adopt the best-fit value. 
Clearly, other reasonable assumptions will lead to somewhat different
$\omb h^2$, and moreover the result (or at least
the error budget) will certainly change as additional WMAP data
becomes available.  
To illustrate this point, 
we will use the WMAP-only results at the end of \S \ref{sect:bbn+cmba}
to illustrate the impact of other assumptions.
Despite these issues, our point in this paper is to illustrate the impact of
the current WMAP results on BBN, and to highlight new
opportunities and challenges for BBN.

\begin{figure}
\hspace*{1.2cm}
\epsfig{file=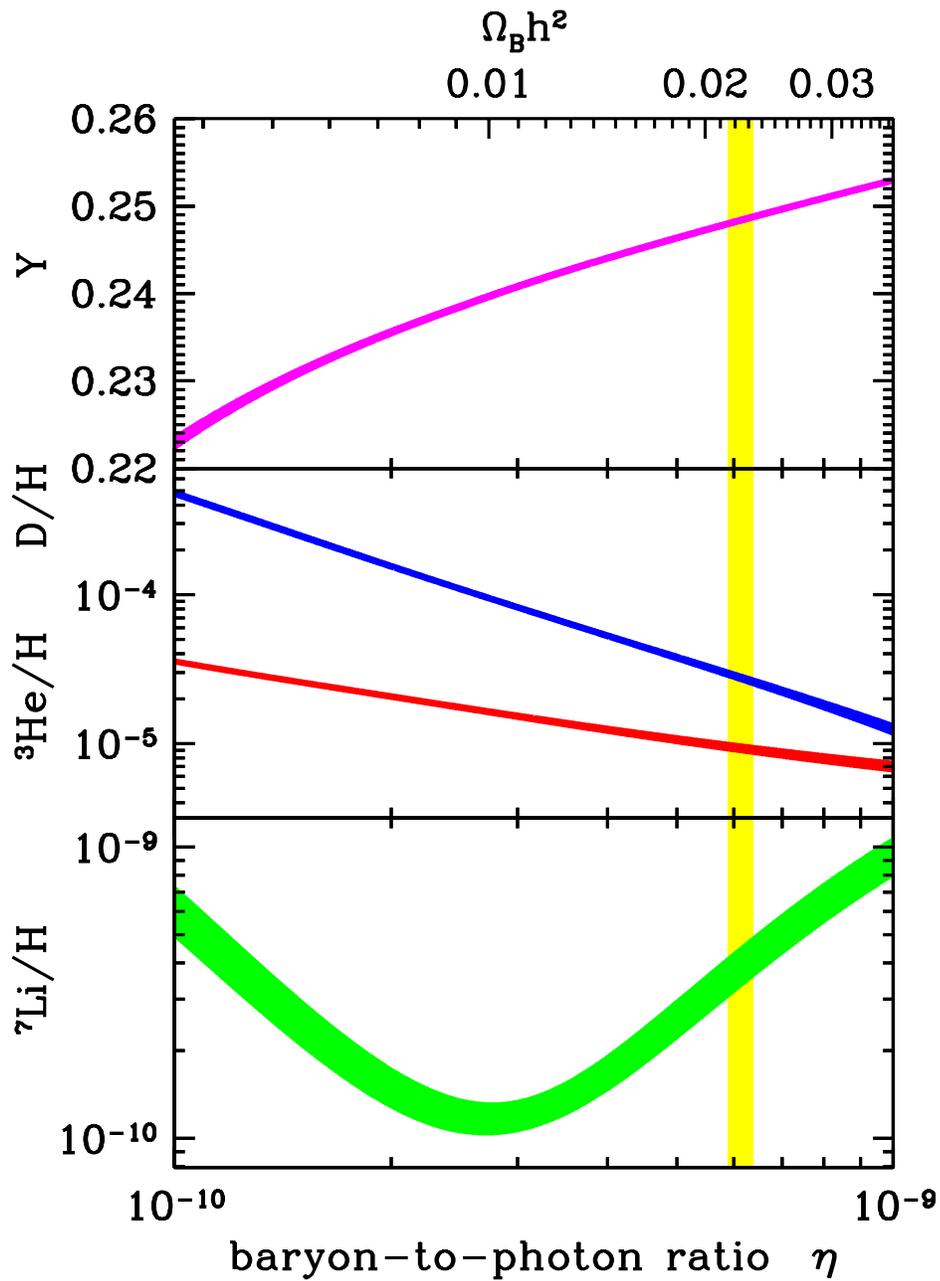, height=7.5in}
\caption{Abundance predictions for standard BBN \protect\cite{cfo1};
the width of the curves give the $1-\sigma$ error range.
The WMAP $\eta$ range (eq.\ \ref{eq:eta-cmb})
is shown in the vertical (yellow) band.}
\label{fig:abvseta-MAP}
\end{figure}

Fig.~\ref{fig:abvseta-MAP} shows
the light element abundance predictions of standard BBN taken from the
recent analysis of \cite{cfo1}, as well as the $\eta$ range determined by
the CMB in eq.\ (\ref{eq:eta-cmb}). 
This range in $\eta$ overlaps with the BBN predicted range (particularly
for the range obtained using D/H) indicating consistency between the BBN
and CMB determinations of $\eta$.
These two techniques involve very different physics,
at different epochs, and rely on observations with
completely different systematics.
Thus, these are independent measurements of the cosmic
baryon content, and their agreement signals that 
the standard hot big bang cosmology has passed a crucial
test in impressive fashion.  

However, we recall that the BBN $\eta$ range based on \li7 and \he4 are in
poor agreement with D. This internal tension to BBN also guarantees that
at least {\em one} element must disagree with the CMB.  However, now the
CMB can act as a ``tiebreaker,'' strongly suggesting that the D/H
measurements are accurate, while both the \he4 and \li7 abundances are
systematically small. This is just one example of the new kinds of
analysis now made possible by using the high-precision CMB $\eta$ as an
{\em input} to BBN \cite{cfo2}. We now turn to a survey of other such
possibilities.

\subsection{Using BBN and the CMB to Probe Astrophysics}
\label{sect:bbn+cmba}

\begin{figure}
\hspace*{1.2cm}
\epsfig{file=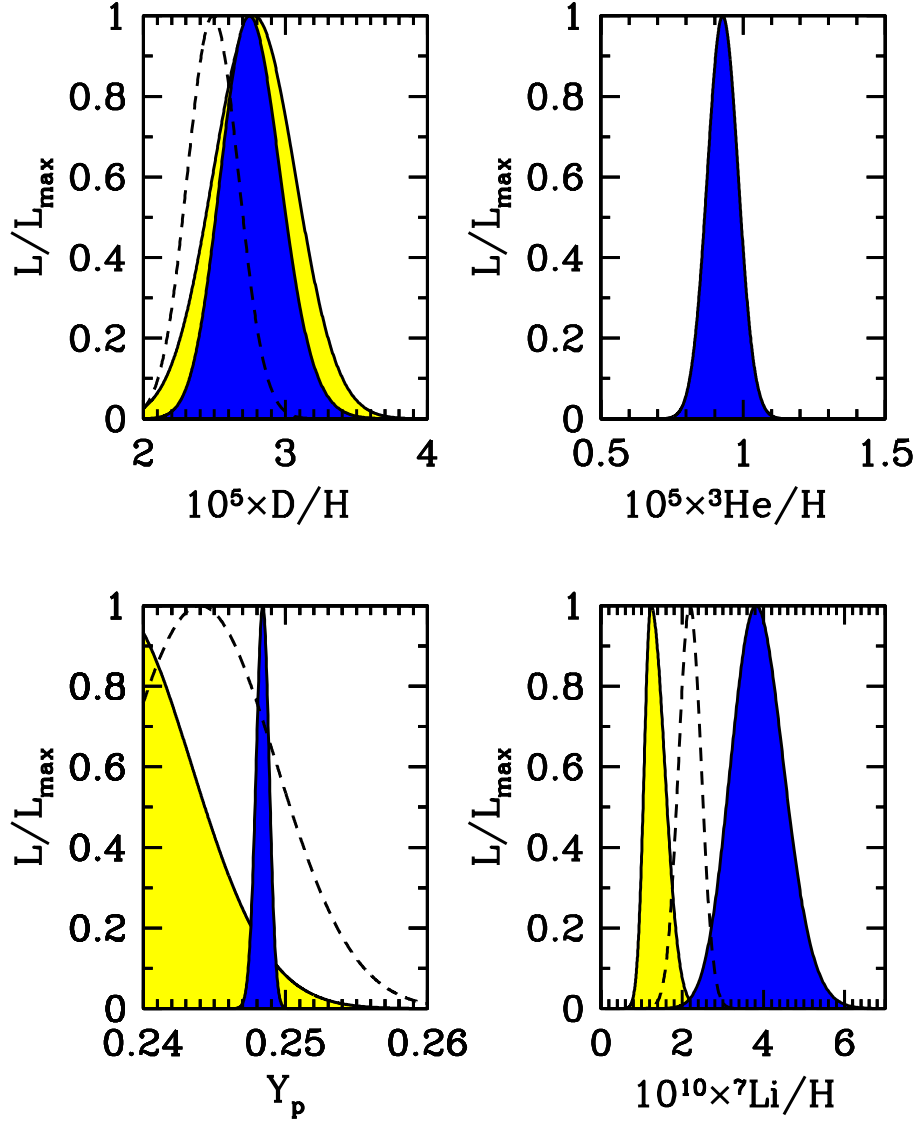, height=7.10in}
\vspace*{-1.0cm}
\caption{
\label{fig:abs-MAP}
Primordial light element abundances as predicted by BBN and
WMAP (dark shaded regions).  
Different observational assessments of primordial abundances 
are plotted as follows:
{\bf (a)} the light shaded
region shows ${\rm D/H} = (2.78 \pm 0.29) \times 10^{-5}$ 
\cite{bt}-\cite{pet},
while the dashed curve shows ${\rm D/H} = (2.49 \pm 0.18) \times 10^{-5}$ 
\cite{omeara,kirkman};
{\bf (b)} no observations plotted
{\bf (c)} the light shaded region shows 
$Y_p = 0.238 \pm 0.002 \pm 0.005$ \cite{fo98},
while the dashed curve shows 
$Y_p = 0.244 \pm 0.002 \pm 0.005$ \cite{iz};
{\bf (d)} the light shaded region shows
\li7/H = $1.23^{+0.34}_{-0.16} \times 10^{-10}$ \cite{ryan},
while the dashed curve shows  
\li7/H = $(2.19\pm 0.28) \times 10^{-10}$ \cite{bon}.
}
\end{figure}

In light of the WMAP determination of $\eta$ (eq. \ref{eq:eta-cmb}),
we now have a very precise prediction for the primordial abundances of
all of the light elements. Our new BBN predictions for each of the
light element abundances are shown in Fig. \ref{fig:abs-MAP} by the
dark shaded distributions. When these are compared to the
observational abundances (shown as the lighter shaded distributions)
the most conservative interpretation of any discrepancy is a
systematic effect in observational determination.  These differences
offer a unique window into the astrophysical processes which are
related to the abundance measurement in both primitive and evolved
systems. We describe each of these
briefly in turn. 

The primordial D/H abundance is predicted to be:
\beq
({\rm D/H})_p = 2.75^{+0.24}_{-0.19} \times 10^{-5}
\label{dpred}
\eeq
a precision of about 8\%.\footnote{
Note here and throughout that the uncertainties quoted
are at the $1\sigma$ or 68\% central confidence limit,
unless otherwise noted.}
For comparison, the uncertainty in the BBN prediction alone
at this $\eta$ is about 4\%, so that the CMB error in $\eta$ dominates,
but as this improves the BBN error will become significant unless
it is reduced.
We note that the predicted value in eq. \ref{dpred} is slightly higher than 
the value of D/H = $2.62^{+0.18}_{-0.20} \times 10^{-5}$ quoted in \cite{wmap},
this is largely due to our use of the most recent nuclear rates
as determined by the NACRE collaboration \cite{nacre}; at higher values of eta,
this leads to 5--10\% more D/H than older rates \cite{cfo2}.
As one can see from Fig.\ \ref{fig:abs-MAP}a, this is in excellent
agreement with the average of the 5 best determined quasar absorption system 
abundances \cite{bt,omeara,kirkman,pet}
which give 
${\rm D/H} = (2.78 \pm 0.29) \times 10^{-5}$.  
It appears that deuterium
in the two systems with multiple-line measurements \cite{omeara,kirkman},
with ${\rm D/H} = (2.49 \pm 0.18) \times 10^{-5}$,
may be systematically low 
(as are the DLA systems in general \cite{omeara,kirkman,pet}); 
however, 
it may be that the error budget is underestimated \cite{kirkman}.

When taken in conjunction with local ISM determinations of
D/H, we see that D/H has been destroyed by only a factor of $\la 2$,
which further implies that the galactic evolution in the disk of our
Galaxy has been rather tame compared with the degree of cosmic evolution
as evidenced by the cosmic star formation rate (see, e.g.~\cite{cova}).  In
fact, we can quantify the fraction of local material that has passed
through stars:  adopting the recent FUSE Local Bubble value of $({\rm
D/H})_{\rm ISM} = (1.52 \pm 0.08) \times 10^{-5}$ \cite{moos}, we see
that $D_{\rm ISM}/D_p = 55^{+6}_{-4}$\% of the Local Bubble material has
never passed through a star.  The FUSE data strongly suggests that D/H
varies outside of the Local Bubble, so that the $D/D_p$ ratios measure
the unprocessed fraction towards each line of sight sampled.

The \he3 abundance is predicted to be:
\beq
{\rm \he3/H} = 9.28^{+0.55}_{-0.54} \times 10^{-6}
\eeq
Unfortunately, as has been argued repeatedly, it is very difficult to 
use local \he3 abundance measurements in conjunction with the BBN value.
The primary reason is our uncertainty in the stellar and chemical
evolution of this isotope over the history of our
Galaxy. Nevertheless, some general statements can be made. For the
most part, the average \he3 abundance seen in Galactic \hii\ regions
\cite{brb} is slightly higher than the above primordial value although the
uncertainties are large.  A few of the systems show abundances at or
below this, while most lie above. Thus one may be tempted to conclude
that, averaged over initial masses,  stars are net producers of \he3. On
the other hand, if the \hii\ regions with abundances apparently {\em
below} the primordial level can be confirmed to be \he3-poor, this
would underscore the difficulty of using \he3 to do cosmology, but
would at the same time offer important hints into low-mass stellar 
evolution as well as the chemical evolution of the Galaxy and its
\hii\ regions \cite{vofc}.

The \he4 abundance is predicted to be:
\beq
Y_p = 0.2484^{+0.0004}_{-0.0005}
\eeq
This value is considerably higher than any prior determination of the 
primordial \he4 abundance.  Indeed it is higher than well over half of
the over 70 low metallicity \hii\ region determinations
\cite{iz,oss,fo98,peim}. While it has been recognized that there are
important systematic effects which have been underestimated \cite{OSk}, it
was believed (or at least hoped) that not all of the \hii\ regions
suffered from these.  Among the most probable cause for a  serious
underestimate of the \he4 abundance is underlying stellar absorption.
Whether or not this effect can account for the serious discrepancy
now uncovered remains to be seen.  Note that the `observed'  distribution
shown in Fig.~\ref{fig:abs-MAP}c  already includes an estimate of the
likely systematic uncertainties.

The \li7 abundance is predicted to be:
\beq
{\rm \li7/H} = 3.82^{+0.73}_{-0.60} \times 10^{-10}
\label{li7}
\eeq
This value is in clear contradiction with most estimates of the 
primordial Li abundance. The question of systematic uncertainties is
now a serious and pressing issue. A thorough discussion of possible
systematic uncertainties was presented in \cite{ryan}.  The result of
that analyses was a \li7 abundance of \li7/H = $1.23^{+0.34}_{-0.16}
\times 10^{-10}$ which is a factor of 3 below the WMAP value, and
almost a factor of 2 below even when systematics are stretched to
maximize the \li7 abundance.  Once again, the most conservative
conclusion that one can reach is that the systematic uncertainties
have been underestimated. One possible culprit in the case of \li7 is
the assumed set of stellar parameters needed to extract an atmospheric
abundance.  In particular, the abundance is very sensitive to
the adopted surface temperature which itself is derived from
other stellar observables. However, even a recent study \cite{bon}
with temperatures based on H$\alpha$ lines (considered to give 
systematically high temperatures) yields \li7/H = $(2.19\pm 0.28) \times
10^{-10}$. Another often discussed possibility is the depletion of
atmospheric \li7.  This possibility faces the strong constraint that
the observed lithium abundances show extremely little
dispersion, making it unlikely that stellar processes which depend on
the temperature, mass, and rotation velocity of the star
all destroy \li7 by the same amount.  To be sure, uniform
depletion factors of order 0.2 dex (a factor of 1.6) have been 
discussed \cite{dep}. It is clear that either (or both) the base-line
abundances of
\li7 have been poorly derived or stellar depletion is far more important than 
previously thought.  Of course, it is possible that if systematic
errors can be ruled out, a persistent discrepancy in \li7 could
point to new physics.

We also note that the WMAP determination, eq.\ (\ref{li7}), has 
important implications for Galactic cosmic-ray nucleosynthesis
(GCRN).  A non-negligible component of \li7 is produced together with
\li6 by  GCRN, predominantly from $\alpha+\alpha$ fusion 
\cite{li6}. 
Since this process is
the only known source of \li6, and the abundance of \li6 is determined 
as the ratio \li6/\li7 in the same metal poor stars, the enhanced
primordial \li7 abundance also implies more GCRN than previously thought.
This in turn has important
implications for cosmic rays in the proto-Galaxy.

As noted in \S \ref{sect:cmb},
the baryon density derived from WMAP depends on the 
assumptions--choices of priors and non-WMAP data--which
enter into the analysis.
Among the suite of models presented by WMAP,
the variations in $\omb h^2$ span a range
of approximately $2\sigma$. Thus, it is of interest
to see the impact of other choices.
As noted above, the baryon density we have adopted (eqs.\ \ref{eq:eta-cmb})
comes from the WMAP ``best-fit'' model, which includes
other data sets which have small but statistically significant
effects on the inferred baryon density.
We thus use the WMAP-only data results to illustrate
the effect of the other data sets on the results.

The WMAP-only baryon density results \cite{wmap} for
a constant primordial spectral index gives
$\omb = 0.024 \pm 0.001$, or $\eta_{10} = 6.58 \pm 0.27$.
Using these values and BBN theory we find
${\rm D/H} = 2.47^{+0.22}_{-0.18} \times 10^{-5}$,
${\rm \he3/H} = 8.89^{+0.55}_{-0.53} \times 10^{-6}$,
$Y_p = 0.2491^{+0.004}_{-0.005}$,
and 
${\rm \li7/H} = 4.39^{+0.83}_{-0.69} \times 10^{-10}$.
We see that the lower D/H value is still in good
agreement with the world average, and
actually in better agreement with the
two best systems.
Both \he4 and \li7 are pushed somewhat further from the
observed levels we have adopted, further pointing
to systematic errors (or possibly new physics).
Thus, while the quantitative differences are significant,
the qualitative conclusions of this section remain the same.

\subsection{Using BBN and the CMB to Probe Particle Physics}
\label{sect:bbn+cmbp}

\begin{figure}
\hspace*{1.2cm}
\epsfig{file=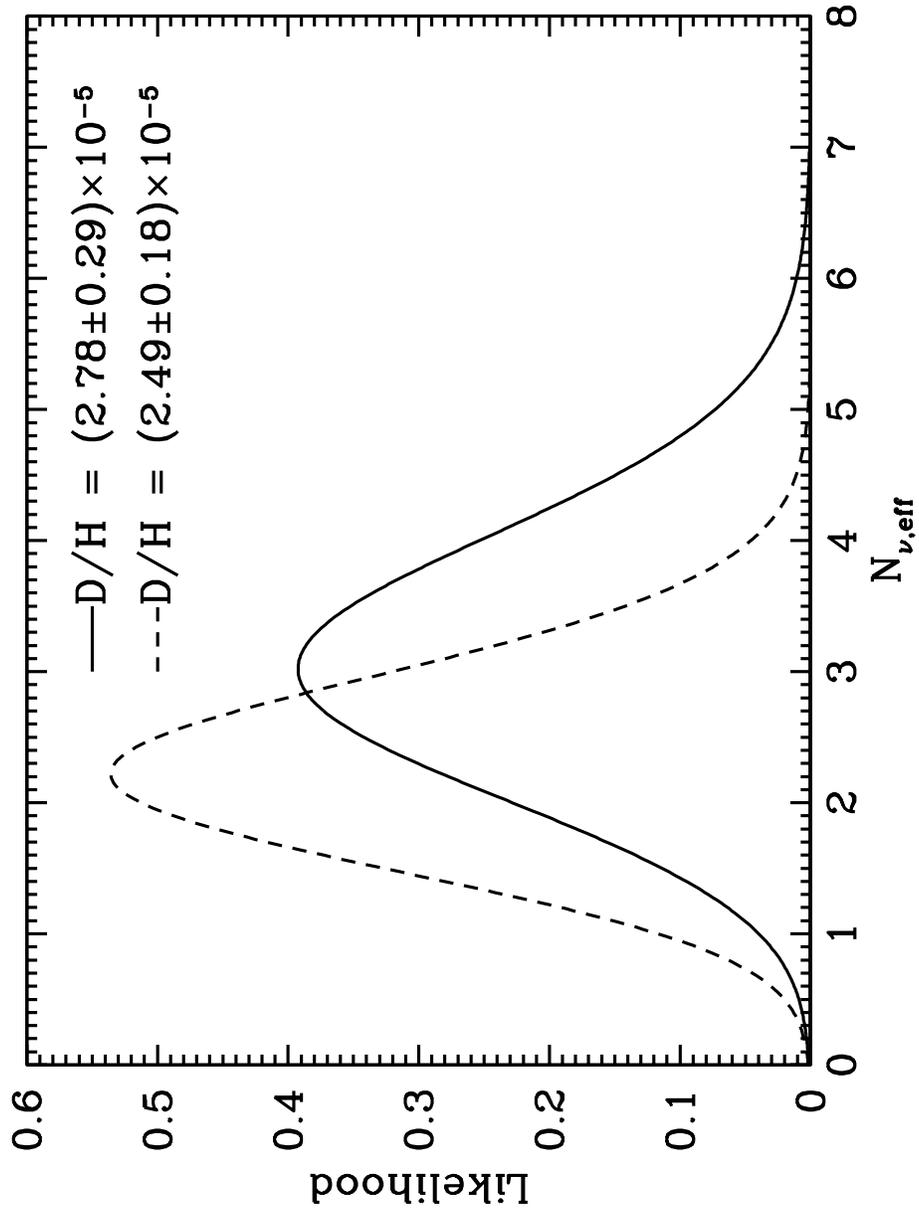, height=7.5in}
\vspace*{-1.0cm}
\caption{
\label{fig:N_nu}
Likelihoods for $\nnu$ as predicted by the WMAP 
$\eta$ (eq.\ \ref{eq:eta-cmb}) and light element observations as in Fig.\ \ref{fig:abs-MAP}.
}
\end{figure}

With the goal of maintaining concordance, we examine how
sharply we can deviate from the standard model.  Often the effect of new
physics can be parameterized in terms of additional relativistic
degrees of freedom, usually expressed in terms of the effective number of
neutrino species $\nnu$, with standard BBN having
$\nnu = 3$.  Traditionally, D or \li7
observations were used to fix the baryon density and the \he4 mass
fraction, was used to fix $\nnu$.
These limits are thoroughly described elsewhere \cite{oldnnu,sarkar,fior}.  
Moreover, as we have noted, the observed \he4 appears lower
than the WMAP+BBN value.
This discrepancy likely is due to systematic errors
(but could point to new physics).
Until this situation is better understood, 
caution is in order. 
Fortunately, in the post-WMAP era, we
can now use the CMB-determined baryon density (eq.~\ref{eq:eta-cmb}), to
remove it as a free parameter from BBN theory and use any or all
abundance observations to constrain $\nnu$ \cite{cfo2}. 
In particular, we have computed the likelihood distributions for $\nnu$
using $\eta_{\rm CMB}$ from WMAP and different observations 
of the primordial D abundances; the results appear
in Fig.~\ref{fig:N_nu}.\footnote{Note that we have neglected
the CMB's own sensitivity to $\nnu$; since the
CMB values for $\eta$ and $\nnu$ are essentially independent
\cite{hann,kssg,crot}, this does not bias our results, but 
means that ours is a more conservative limit.   }

Unlike \he4, deuterium does not appear to suffer from large
systematics.  It is simply limited by the low number statistics
due to the difficulty of finding high-redshift systems well-suited
for accurate D/H determinations.  Given
that D predictions from WMAP agree quite well with observations, we
can now use D to place an interesting limit on $\nnu$.  
D is not as sensitive to $\nnu$ as
\he4 is, but nonetheless it does have a significant dependence.
The relative error in the observed abundance of D/H ranges from 7-10\%,
depending on what systems are chosen for averaging.  If the five most
reliable systems are chosen, the peak of the $\nnu$ likelihood
distribution lies at $\nnu \approx 3.0$, with a width of $\Delta\nnu
\approx 1.0$ as seen in Fig.~\ref{fig:N_nu}.  However, if we limit our
sample to the two D systems that have had multiple absorption features
observed, then the peak shifts to $\nnu \approx 2.2$, with a width of
$\Delta\nnu \approx 0.7$.  Given the low number of observations, it is
difficult to qualify these results.  The differences could be statistical
in nature, or could be hinting at some underlying systematic affecting
these systems.  Adopting the five system D average, D/H$ = (2.78 \pm
0.29)\times 10^{-5}$, we get the following constraints on $\nnu$:
\beqar
&\hat{N}_{\rm \nu,eff}&\!\!\!\! = 3.02 \\
2.10 <\!\!\!\! &\nnu&\!\!\!\! < 4.14 \ \ \ \mbox{(68\% CCL)} \nonumber\\
1.26 <\!\!\!\! &\nnu&\!\!\!\! < 5.22 \ \ \ \mbox{(95\% CCL)} \nonumber
\eeqar
where CCL is central confidence limit.
Using a standard model prior assuming $\nnu \ge 3.0$
\cite{prior}, the corresponding 95\% CL upper limits  are: 
 $\nnu < 5.19$ for D/H = 2.78 $\times 10^{-5}$;
$\nnu < 4.20$ for D/H = 2.49 $\times 10^{-5}$.
For comparison, we also quote the corresponding limits based on
\he4: $\nnu < 3.40$ for $Y_P = 0.238$; $\nnu < 3.64$ for $Y_P
= 0.244$ also assuming the prior of $\nnu > 3.0$.
Also for comparison, we note that 
note that the CMB itself also constrains
$\nnu$ \cite{hann,kssg,crot}.  From the WMAP data alone, $\nnu < 6$ (95\% CL) \cite{crot}.  
Note that it is conceivable that an evolving nonstandard
component could lead to different $\nnu$ at the BBN and CMB 
epochs; as the data improve, this could be tested.

The new power of D to probe early universe physics will grow
with the increasing precision in $\eta_{\rm CMB}$ and particularly
with increasing accuracy in observed D/H.
A 3\% measurement in D will allow it to
become the dominant constraint on $\nnu$ \cite{cfo2}.

\section{Discussion and Conclusions}

\label{sect:discuss}

Primordial nucleosynthesis has entered a new era.
With the precision observations of WMAP,
the CMB has become the premier cosmic baryometer.
The independent BBN and CMB predictions
for $\eta$ are in good agreement (particularly when D is used in BBN),
indicating that cosmology has passed a fundamental test.
Moreover, this agreement allows us to use BBN in a new way, as
the CMB removes $\eta$ as a free parameter.
One can then adopt the standard BBN predictions,
and use $\eta_{\rm CMB}$ to infer primordial abundances;
by comparing these to light element abundances in
different settings, one gains new insight into
the astrophysics of stars, \hii\ regions, cosmic rays,
and chemical evolution, to name a few examples.
Alternately, WMAP transforms BBN into a sharper
probe of new physics in the early universe;
with $\eta_{\rm CMB}$ fixed, {\em all} of the light elements
constrain non-standard nucleosynthesis, with
$\nnu$ being one example.

As BBN assumes a new role, much work remains to be done.
To leverage the power of the WMAP precision
requires the highest possible precision in light element
observations.
Further improvements in the primordial D abundance
can open the door to D as a powerful probe of early universe physics.
Improved \he3 observations can offer new insight into
stellar and chemical evolution.
And perhaps most pressing, the WMAP prediction for primordial
\he4 and particularly \li7 are higher than
the current observed abundances; 
it remains to be resolved what systematic effects
(or new physics!) has led to this discrepancy.

WMAP also demands improvements in BBN theory.
While the basic calculation is sound, 
accuracy of the WMAP light element predictions (Fig.\ \ref{fig:abs-MAP})
is or soon will be limited by the errors in BBN theory.
These in turn arise from uncertainty in nuclear reaction 
cross sections \cite{nuc,cfo1}.
In particular, the \li7 prediction is completely dominated
by the nuclear errors, especially that in
the $\he3(\alpha,\gamma)\be7$ reaction.
The error in \he3 is also due to BBN uncertainties,
in this case the $d(p,\gamma)\he3$ and
$\he3(d,p)\he4$ reactions dominate the uncertainty.
About half of the uncertainty in the CMB + BBN prediction of D is due to
BBN errors, where again $d(p,\gamma)\he3$ is important, as
well as $p(n,\gamma)d$ and $d(d,n)\he3$.
We encourage intensified efforts to obtain high-precision 
measurements of these reactions, and their uncertainties.

In closing, it is impressive that our
now-exquisite understanding of the universe
at $z \sim 1000$ 
also confirms our understanding of the
universe at $z \sim 10^{10}$.
This agreement lends
great confidence in the soundness of the hot big bang cosmology,
and impels our search deeper into the early universe.

{\bf Acknowledgments}

We thank Benjamin Wandelt for helpful conversations,
and Ken Nollett for useful discussions regarding the differences
between our BBN predictions.
The work of K.A.O. was partially supported by DOE grant
DE--FG02--94ER--40823.
The work of B.D.F. and R.H.C. 
was supported by the National Science
Foundation under grant AST-0092939.


\begin{thebibliography}{0}


\bibitem{bbn} T. P. Walker, G. Steigman, D. N. Schramm, 
 K. A. Olive and K. Kang, {\it Ap.J.} {\bf 376} (1991) 51;
K. A. Olive, G. Steigman, and T. P. Walker, {\it Phys. Rep.} {\bf 333} (2000) 389; 
B. D. Fields and S. Sarkar, 
{\it Phys.\ Rev.} {\bf D66} (2002) 010001.

\bibitem{sarkar}
S. Sarkar, {\it Rep. Prog. Phys.} {\bf 59} (1996) 1493.

\bibitem{st}
D.~N.~Schramm and M.~S.~Turner,
Rev.\ Mod.\ Phys.\  {\bf 70}, 303 (1998)
[arXiv:astro-ph/9706069].

\bibitem{wmap}
C.L.~Bennett {\it et al.},
{\it Astrophys.~J.} submitted,
arXiv:astro-ph/0302207;
D.~N.~Spergel {\it et al.},
{\it Astrophys.~J.} submitted,
arXiv:astro-ph/0302209;


\bibitem{cfo2}
R.~H.~Cyburt, B.~D.~Fields and K.~A.~Olive,
Astropart.\ Phys.\  {\bf 17} (2002) 87
[arXiv:astro-ph/0105397].

\bibitem{kssg}
J.~P.~Kneller, R.~J.~Scherrer, G.~Steigman and T.~P.~Walker,
Phys.\ Rev.\ D {\bf 64}, 123506 (2001)
[arXiv:astro-ph/0101386];
S.~H.~Hansen, G.~Mangano, A.~Melchiorri, G.~Miele and O.~Pisanti,
Phys.\ Rev.\ D {\bf 65}, 023511 (2002)
[arXiv:astro-ph/0105385];
P.~Di Bari and R.~Foot,
Phys.\ Rev.\ D {\bf 63}, 043008 (2001)
[arXiv:hep-ph/0008258].


\bibitem{rafs} H. Reeves, J. Audouze, W. Fowler, and D. N. Schramm,  Ap.
J. {\bf 179} (1976) 909.

\bibitem{ytsso}J.~M.~Yang, M.~S.~Turner, G.~Steigman, D.~N.~Schramm and K.~A.~Olive,
Ap. J.  {\bf 281} (1984) 493.

\bibitem{fo} B.D. Fields and K.A. Olive, {\it Phys. Lett.} {\bf B368}
(1996) 103; \\ B.D. Fields, K. Kainulainen, D. Thomas, and K.A. Olive,
{\it New Astronomy} {\bf 1} (1996) 77.


\bibitem{lik}
N.~Hata, R.~J.~Scherrer, G.~Steigman, D.~Thomas, T.~P.~Walker, S.~Bludman and P.~Langacker,
Phys.\ Rev.\ Lett.\  {\bf 75}, 3977 (1995)
[arXiv:hep-ph/9505319];
N.~Hata, G.~Steigman, S.~Bludman and P.~Langacker,
Phys.\ Rev.\ D {\bf 55}, 540 (1997)
[arXiv:astro-ph/9603087];
S.~Esposito, G.~Mangano, G.~Miele and O.~Pisanti,
Nucl.\ Phys.\ B {\bf 568} (2000) 421
[arXiv:astro-ph/9906232];
S.~Burles, K.~M.~Nollett and M.~S.~Turner,
Astrophys.\ J.\  {\bf 552}, L1 (2001)
[arXiv:astro-ph/0010171].

\bibitem{fior}
G.~Fiorentini, E.~Lisi, S.~Sarkar and F.~L.~Villante,
Phys.\ Rev.\ D {\bf 58}, 063506 (1998)
[arXiv:astro-ph/9803177];


\bibitem{cfo1}
R.~H.~Cyburt, B.~D.~Fields and K.~A.~Olive,
New Astron.\  {\bf 6} (1996) 215
[arXiv:astro-ph/0102179].

\bibitem{brb}
T.~M.~Bania, R.~T.~Rood and D.~S.~Balser, 
{\it Nature}  {\bf 415} (2002) 54.

\bibitem{vofc}
E.~Vangioni-Flam, K.~A.~Olive, B.~D.~Fields and M.~Casse,
arXiv:astro-ph/0207583.

\bibitem{zss}
M.~Zaldarriaga, D.~N.~Spergel and U.~Seljak,
Ap. J. {\bf 488} (1997) 1
[arXiv:astro-ph/9702157];
J.~R.~Bond, G.~Efstathiou and M.~Tegmark,
Mon.\ Not.\ Roy.\ Astron.\ Soc.\  {\bf 291} (1997) L33
[arXiv:astro-ph/9702100];
G.~Jungman, M.~Kamionkowski, A.~Kosowsky and D.~N.~Spergel,
Phys.\ Rev.\ D {\bf 54} (1996) 1332
[arXiv:astro-ph/9512139].

\bibitem{oldcmb} R.~Stompor {\it et al.},
{\it Ap.J.}  {\bf 561} (2001) L7
[arXiv:astro-ph/0105062];
C.~B.~Netterfield {\it et al.} 
 [Boomerang Collaboration], {\it Ap. J.} {\bf 571} (2002) 604
[arXiv:astro-ph/0104460];
C.~Pryke, N.~W.~Halverson, E.~M.~Leitch, J.~Kovac, J.~E.~Carlstrom,
W.~L.~Holzapfel and M.~Dragovan, 
{\it Ap. J.} {\bf 568} (2002) 46
[arXiv:astro-ph/0104490].

\bibitem{cbi}
J.~L.~Sievers {\it et al.},
arXiv:astro-ph/0205387.

\bibitem{acb}
J.~H.~Goldstein {\it et al.},
arXiv:astro-ph/0212517.

\bibitem{2df}
W.~J.~Percival  {\it et al.},
{\it Monthly Not.~Royal Astr.~Soc.},
{\bf 337}, 1297 (2001).

\bibitem{bt} S. Burles and D. Tytler, {\it Ap.J.} {\bf 499}, 699 (1998);
{\it Ap.J.} {\bf 507}, 732 (1998).

\bibitem{omeara}
J.~M.~O'Meara, D.~Tytler, D.~Kirkman, N.~Suzuki, J.~X.~Prochaska, D.~Lubin and A.~M.~Wolfe,
Astrophys.\ J.\  {\bf 552}, 718 (2001)
[arXiv:astro-ph/0011179].

\bibitem{kirkman}
D.~Kirkman, D.~Tytler, N.~Suzuki, J.~M.~O'Meara and D.~Lubin,
arXiv:astro-ph/0302006.

\bibitem{pet}
M.~Pettini and D.~V.~Bowen,
Astrophys.\ J.\  {\bf 560}, 41 (2001)
[arXiv:astro-ph/0104474].

\bibitem{nacre}
C.~Angulo et al.~(NACRE Collaboration), 
{\it Nuc.\ Phys.\ A} {\bf 656} (1999) 3.

\bibitem{fo98}
B.~D.~Fields and K.~A.~Olive,
{\it Ap.\ J.} {\bf 506} (1998) 177
[arXiv:astro-ph/9803297].

\bibitem{iz}
Y.~I.~Izotov and T.~X.~Thuan,
Ap.\ J. {\bf 500} (1998) 188;
Y.~I.~Izotov, T.~X.~Thuan, and V.~A.~Lipovetsky,
Ap.\ J. {\bf 108} (1997) 1;
Y.~I.~Izotov, T.~X.~Thuan, and V.~A.~Lipovetsky,
Ap.\ J. {\bf 435} (1994) 647.

\bibitem{ryan} S.G. Ryan,  T.C. Beers, K.A. Olive, B.D. Fields, and J.E.
Norris, {\it Ap.J. Lett.} {\bf 530}, L57 (2000).

\bibitem{bon}
P.~Bonifacio, et al., 
Astron.~Astrophys., {\bf 390} (2002) 91.



\bibitem{cova}
M.~Casse, K.~A.~Olive, E.~Vangioni-Flam and J.~Audouze,
New Ast. {\bf 3} (1998) 259
[arXiv:astro-ph/9712261].

\bibitem{moos} H.W. Moos, et al.\ 
Ap. J. Supp. {\bf 140} (2002) 3. 



\bibitem{oss}
K.A. Olive, E. Skillman, and G. Steigman, Ap.J.
{\bf 483} (1997) 788.



\bibitem{peim} 
M. Peimbert, A. Peimbert and M. T. Ruiz,
Ap.\ J. {\bf 541} (2000) 688;
A. Peimbert, M. Peimbert and V. Luridiana, Ap.\ J. {\bf 565} (2002) 668.


\bibitem{OSk} K.A. Olive, and E. Skillman, {\it New Ast.}
{\bf 6} (2001) 119
[arXiv:astro-ph/0007081];
D. Sauer, and K. Jedamzik, {\it A.A.} {\bf 381}, 361 (2002);
R.~Gruenwald, G.~Steigman and S.~M.~Viegas,
{\em Astrophys.~J.} {\bf 567} (2002) 931.
[arXiv:astro-ph/0109071].


\bibitem{dep}
S. Vauclair,and C. Charbonnel, Ap. J. {\bf 502} (1998) 372;
M.~H.~Pinsonneault, T.~P.~Walker, G.~Steigman and V.~K.~Narayanan,
Ap. J. {\bf 527} (1998) 180
[arXiv:astro-ph/9803073];
M.~H.~Pinsonneault, G.~Steigman, T.~P.~Walker, and V.~K.~Narayanan,
Ap. J. {\bf 574} (2002) 398
arXiv:astro-ph/0105439.

\bibitem{li6}
E. Vangioni-Flam, et al., New Astron.~{\bf 4} (1999) 245;
B.D. Fields and K.A. Olive,
New Astron.~{\bf 4} (1999) 255;
T.P.~Walker and G. Steigman, Ap. J. Lett.~{\bf 385} (1991) 13;
T.~Montmerle, Ap.~J. {\bf 217} (1977) 878.


\bibitem{oldnnu}
G. Steigman, D.N. Schramm, and J. Gunn,
   Phys. Lett. {\bf B66} (1977) 202;
K.A. Olive and D. Thomas, Astropart. Phys. {\bf 11} (1999)
403 (1999).  

\bibitem{prior}
 K.~A.~Olive and G.~Steigman,
Phys.\ Lett.\ B {\bf 354} (1995) 357.

\bibitem{hann}
S.~Hannestad,
Phys.\ Rev.\ Lett.\  {\bf 85}, 4203 (2000)
[arXiv:astro-ph/0005018].

\bibitem{crot}
P.~Crotty, J. Lesgourgues, and S.~Pastor,
arXiv:astro-ph/0302337.

\bibitem{nuc}
M. Smith, L. Kawano, and R.A. Malaney,  {\it Ap.J. Supp.}
{\bf 85}, 219 (1993);
A.~Coc, E.~Vangioni-Flam, M.~Casse and M.~Rabiet,
Phys.\ Rev.\ D {\bf 65}, 043510 (2002);
K.~M.~Nollett and S.~Burles,
{\it Phys.\ Rev.}  {\bf D61} (2000) 123505
[arXiv:astro-ph/0001440].


\end{thebibliography}
\end{document}